\documentclass[superscriptaddress,floatfix,pra,twocolumn,amsmath,amssymb]{revtex4-2}
\usepackage{caption}
\usepackage{subcaption}
\usepackage{float}
\usepackage{lipsum}
\usepackage{graphicx}
\usepackage{xcolor}
\usepackage{mathrsfs}
\usepackage{mathtools}
\usepackage{braket}
\usepackage{multirow}
\usepackage{algorithm}  
\usepackage{algpseudocode}
\usepackage{placeins}
\usepackage{enumitem}  
\usepackage{placeins}
\usepackage{afterpage}
\usepackage{url}
\usepackage{booktabs} 
\usepackage{bm}
\usepackage[margin=1in]{geometry}
\usepackage{tabularx}
\usepackage{array}
\usepackage{adjustbox} 
\usepackage{hyperref} 
\hypersetup{allcolors=blue}

\newcommand{\bracket}[1]{\langle 1\rangle}

\begin{document}

\title{Resource-bounded controllability benchmarking of open quantum systems}
\author{Yule Mayevsky}
\address{Quantum Photonics Laboratory and Centre for Quantum Computation and Communication Technology, RMIT University, Melbourne, VIC 3000, Australia}
\email{yulemayevsky@outlook.com}
\author{Akram Youssry}
\address{Quantum Photonics Laboratory and Centre for Quantum Computation and Communication Technology, RMIT University, Melbourne, VIC 3000, Australia}
\address{School of Electrical Engineering and Telecommunications, UNSW Sydney, Sydney, NSW 2052, Australia}

\author{Alberto Peruzzo}
\address{Quantum Photonics Laboratory and Centre for Quantum Computation and Communication Technology, RMIT University, Melbourne, VIC 3000, Australia}
\address{Quandela, Massy, France}

\begin{abstract}
    \noindent This paper develops a resource-bounded framework for evaluating practical controllability in open quantum systems using a trained graybox response model. Rather than treating controllability as a binary property of an idealised Hamiltonian model, the proposed approach evaluates the best-achievable process fidelity over a finite, hardware-realisable pulse family under explicit control constraints. The graybox model retains the known coherent dynamics while learning control-dependent open-system distortions from pulse--response data. The resulting surrogate predictions are used to reconstruct the implemented processes and compare them with Haar-random target gates. Practical controllability is then characterised through the distribution of best-achievable infidelities and an area-based summary metric. The framework is demonstrated for a driven qubit under closed-system, classical-noise, and combined quantum-plus-classical-noise dynamics, with pulse amplitude and inverse Gaussian width used as the control-resource coordinates. The results show how finite control resources and open-system noise jointly constrain the gate performance attainable by the chosen pulse family.
\end{abstract}
\maketitle
\section{Introduction}
The practical realisation of quantum technologies relies on implementing desired quantum operations with high fidelity under finite hardware resources. In idealised closed-system settings, controllability admits a clean binary characterisation: a finite-dimensional system is fully controllable if the Lie algebra generated by the drift and control Hamiltonians spans $\mathrm{su}(d)$ \cite{dalessandro2007introduction,schirmer2001complete,albertini2001notions,altafini2001controllability}. This Lie-algebraic criterion underpins much of coherent control theory \cite{rabitz2005landscape}.

Real devices operate in open-system regimes where decoherence, drift, bandwidth limits, cross-talk, calibration error, and non-ideal control pulses induce non-unitary dynamics \cite{breuer2002theory,wiseman2009book,koch2022quantum, Motzoi2009, Schutjens2013, Warren1994}. In this setting, the closed-system Lie-algebraic test is not predictive of what can be implemented reliably at high fidelity, particularly when noise is only partially characterized, may depend on the applied control, and may exhibit memory effects \cite{devega2017open, koch2016controlling,arenz2014control,morris2022quantifying, white2020demonstration, giarmatzi2023multi}.

A substantial literature addresses open-system controllability and control, but the resulting notions are typically model and resource-specific, and beyond the closed-system setting the characterisation of reachable open-system dynamics becomes substantially more complicated \cite{dirr2008lietheory,arenz2017lindbladian,dive2015isotropy}.
For Markovian Lindblad dynamics, reachability is naturally formulated in terms of semigroups and analysed using Lie-semigroup/Lie-wedge structures \cite{dirr2008lie}. 
Channel-level perspectives study controllability in the space of CPTP maps, including Kraus-map formulations with ancillas and measurement/feedback resources \cite{wu2007controllability}, and related results emphasise that feedback can enlarge what is achievable compared to open-loop control in dissipative settings \cite{schirmer2003controllability}. 
More recent work also develops constructive open-system control mechanisms where the applied drive modifies the system–environment coupling (control-dependent dissipation), enabling entropy-changing maps and even unitary gate synthesis under dissipation \cite{kallush2022controlling}.

These frameworks are foundational, but they do not directly provide a device-level notion of controllability for a fixed, hardware-realisable control family in the presence of partially unknown open-system noise. 
Accordingly, the contribution of this work is not a new theoretical controllability criterion.
Instead, we treat practical controllability as a resource-bounded device-engineering problem: quantifying how well a fixed admissible pulse family can approximate target operations, and how this capability changes as additional control resources become available. 
This shifts the emphasis from whether a target operation is exactly reachable to the distribution of best-achievable fidelities over a target-gate ensemble.

Evaluating practical controllability in this sense requires scoring many candidate control waveforms by the operation they realise on the device; the bottleneck is therefore an accurate and scalable way to predict device-level gate performance under partially unknown noise \cite{auza2024quantum}. 
Pure physics derived ``whitebox'' models can be systematically inaccurate in high-fidelity regimes due to missing couplings and unmodelled dissipation, while pure data-driven ``blackbox'' models typically require large probing datasets and provide limited diagnostic structure \cite{rabitz2005landscape, Fosel2018}.
We therefore adopt a graybox machine learning model that retains the known dynamics generated by the drift and control Hamiltonians, while learning a compact, control-dependent description of open-system distortions from probing data. Previous applications of graybox learning to single-qubit systems demonstrated high-fidelity gate synthesis under realistic noise, using a mean-squared-error objective between a target gate and the model-predicted gate evolution \cite{youssry2023noise,youssry2022multiaxis,auza2024quantum}. More recently, we showed that this framework can be extended to higher-dimensional systems (qudits) \cite{Mayevsky}. 

In the present work, we build on that in a different direction. Rather than focusing only on gate synthesis, we use the graybox model to study practical open-system controllability under explicit control-resource constraints. We evaluate a discrete control grid, predict the process-level quantities needed for fidelity scoring, and maximise fidelity over admissible controls to obtain the best-achievable performance for each Haar-random target gate. Sweeping a single budget parameter then produces capability curves and area-based controllability summaries that link hardware resources to achievable gate performance. While the framework is dimension-general, we present the numerical controllability study for the concrete case \(d=2\), namely a driven qubit under combined quantum-plus-classical noise.
The paper contributes a resource-bounded formulation of practical controllability in which reachable performance is described by distributions over Haar-random target gates rather than by a binary Lie-algebraic condition. It combines a trained graybox response model with surrogate-based channel reconstruction to evaluate implemented processes over a finite admissible pulse library. The resulting pipeline is then applied to amplitude and bandwidth constraints in a driven qubit model under closed-system, classical-noise, and combined quantum-plus-classical noise regimes, producing both one-dimensional and two-dimensional controllability summaries.

\section{Problem setting}
In this work, we study quantum systems described by a $d$-dimensional Hilbert space. 
The Hamiltonian of the system can be expressed in the form
\begin{align}
H(t) = H_{\text{drift}} + H_{\text{control}}(t)  
 + H_{\text{noise}}(t)   
 \label{equ:Htotal}
\end{align}

We assume the drift and control terms are perfectly known, while the noise term is treated as unknown during control and controllability evaluation. 
This noise term can represent classical stochastic processes, in which case it acts only on the system, or quantum noise, in which case it includes both the bath Hamiltonian and the system--bath interaction \cite{youssry2020characterization, youssry2022multiaxis}. 
The control Hamiltonian $ H_{\text{control}}(t) $ is expressed as a sum of $N_C$ time-dependent control fields $ f_i(t;\vec{\theta}_i) $ parametrised by the vector $\vec{\theta}_i$, modulating the system control operators $ X_i $.
\begin{align}
    H_{\text{control}}(t) = \sum_{i=0}^{N_C} f_i(t;\vec{\theta}_i) X_i  
    \label{equ:Hctrl}
\end{align}

The system evolves from \(t=0\) to \(t=T\) under the Hamiltonian \(H(t)\). The system is initialised in a state \(\rho(0)\), and we assume access to measurements of system observables at the final time. Repeating this over a tomographically complete set of initial states and observables yields the probing data used to train the graybox model.
In prior work \cite{Mayevsky}, we studied the problem of finding control parameters \(\vec{\theta}_i\) that implement a target unitary gate \(G\).
Here, we instead study practical controllability in open-system conditions.
Given a device governed by Eq.~(\ref{equ:Htotal}) with bounded controls of the form Eq.~(\ref{equ:Hctrl}), and subject to an unknown (and potentially non-Markovian) noise term $H_{\mathrm{noise}}(t)$, we ask:

\emph{``Which target operations $G\in\mathrm{SU}(d)$ are achievable to high fidelity on the physical device under explicit hardware constraints, and how does this capability change as additional control resources 
(e.g. amplitude or bandwidth) become available?''}
Accordingly, we study practical controllability under bounded controls in the presence of unknown non-Markovian open system dynamics.

\section{Methodology}

\begin{figure*}[t]
    \centering
    \includegraphics[width=\textwidth]{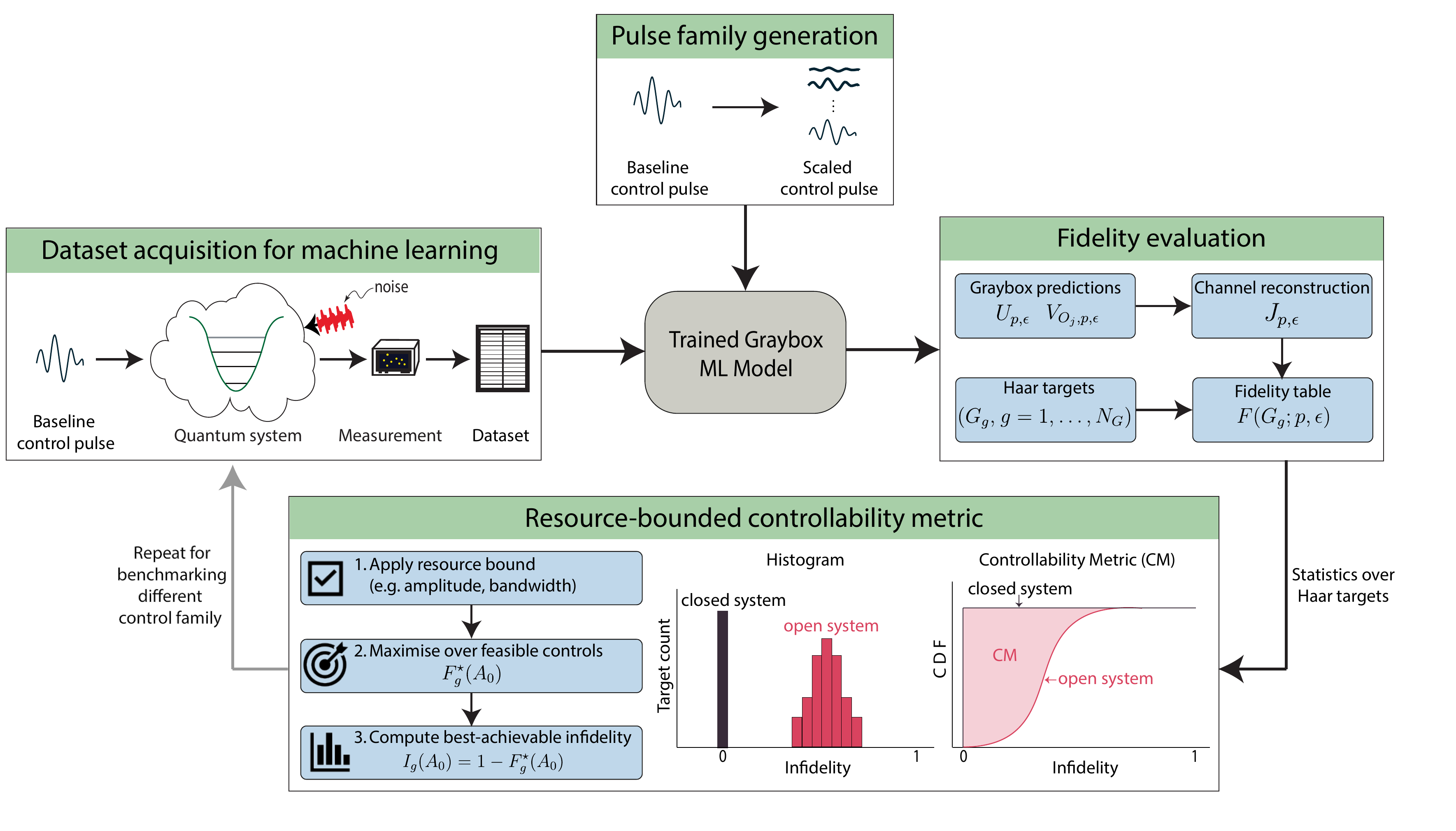}
    \caption{\textbf{Overview of the graybox controllability workflow.} Experimental pulse--response data are first collected and used to train a graybox surrogate model. Each baseline pulse is then expanded into constrained variants by scaling selected pulse parameters such as amplitude or bandwidth. For each variant, the trained graybox predicts the corresponding dynamics, including both the coherent unitary part and the noise contribution, allowing the effective implemented channel to be reconstructed. These implemented channels are then compared against Haar-random target gates through process fidelity. By maximising fidelity over the feasible pulse family for each target, the method builds a statistical picture of practical controllability for open quantum systems under explicit resource constraints.}
    \label{fig:cm_workflow}
\end{figure*}

Starting from a fixed library of hardware-realisable control pulses, we generate constrained pulse families, use the trained graybox model to predict the realised open-system dynamics for each candidate waveform, reconstruct the corresponding quantum channel in a fixed tomography basis, and compare the realised channels against Haar-random target gates using process fidelity.
Practical controllability is then summarised statistically over the admissible subset defined by explicit control budgets.
Figure~\ref{fig:cm_workflow} summarises the controllability evaluation pipeline used throughout this paper.
A central assumption of the method is that the graybox model is valid only over a bounded region of control space. The transformed pulse family is therefore restricted to remain inside the regime on which the surrogate was trained. The global limits used later in the budget construction are not arbitrary; they are inherited from the admissible control range encoded in the training data.
The exact choice of pulse family, the parameter to be scaled, and whether the remaining parameters are held fixed or co-varied is problem dependent. 
Accordingly, the resulting practical-controllability summaries are conditional on the chosen admissible pulse family and its parametrisation.
\subsection{Graybox modelling and prediction}

To evaluate practical controllability under unknown open-system dynamics, we use a graybox model that preserves the known coherent structure while learning an effective representation of the unknown environmental distortion from probing data.
Here, we express the dynamics of the quantum system using the noise operator formalism from \cite{Gerardo,youssry2020characterization}.

We first define the known part of the Hamiltonian,
\begin{align}
H_0(t):=H_{\mathrm{drift}}+H_{\mathrm{control}}(t),
\end{align}
with coherent propagator
\begin{align}
U_0(T)=\mathcal{T}\exp\!\left[-i\int_0^T H_0(s)\,ds\right].
\label{eq:U0_method}
\end{align}
Moving to the interaction picture with respect to $H_0(t)$ gives
\begin{align}
H_I(t)=U_0^\dagger(t)\,H_{\mathrm{noise}}(t)\,U_0(t),
\end{align}
with interaction-picture propagator
\begin{align}
U_I(T)=\mathcal{T}\exp\!\left[-i\int_0^T H_I(s)\,ds\right].
\end{align}
Following the noise-operator formalism, we define
\begin{align}
\widetilde{U}_I(T):=U_0(T)\,U_I(T)\,U_0^\dagger(T).
\end{align}

For an initial system state $\rho(0)$ and a measured system observable $O$ at the final time $T$, the expectation value may be written as
\begin{align}
\langle O(T)\rangle
=
\mathrm{Tr}\!\left[
V_O(T)\,
U_0(T)\,\rho(0)\,U_0^\dagger(T)\,
O
\right],
\label{eq:expectation_noise_operator_method}
\end{align}
where
\begin{align}
V_O(T)
=
\Big\langle
\mathrm{Tr}_B\!\left[
O^{-1}\,\widetilde{U}_I^\dagger(T)\,
O\,
\widetilde{U}_I(T)\,
\rho_B
\right]
\Big\rangle_c .
\label{eq:VO_method}
\end{align}
Here $\rho_B$ is the initial bath state, $\mathrm{Tr}_B(\cdot)$ denotes partial trace over the bath, and $\langle\cdot\rangle_c$ denotes averaging over any classical noise process when present.

The operator $V_O(T)$ captures the control-dependent open-system distortion seen in the measured dynamics. In realistic non-Markovian settings it is generally not available in closed form. We therefore do not attempt to derive it analytically for every candidate waveform. Instead, the graybox surrogate is trained to learn the map
\begin{align}
\vec{\theta}\mapsto \big(U_0(T;\vec{\theta}),\,V_O(T;\vec{\theta})\big),
\end{align}
through measured or simulated probing data, while retaining the known coherent structure explicitly.

\paragraph{Graybox model prediction}
The training data consist of pairs $(\vec{\theta},\vec{E})$, where $\vec{\theta}$ specifies one realisable control waveform and $\vec{E}$ contains the corresponding vector of probing measurement outcomes over an informationally complete basis of initial states and observables. 
The graybox model is trained to predict $\vec{E}$ from $\vec{\theta}$ while internally preserving the known coherent evolution generated by $H_0(t)$ and learning the unknown control-dependent open-system correction through the effective operators $V_O(T)$.

Once training is complete, the model is used as a fast surrogate over a finite evaluation grid of control candidates. Let $p\in\{1,\dots,P\}$ index a baseline pulse drawn from a fixed control library, and let $\epsilon$ index a constrained variant of that pulse. Each pair $(p,\epsilon)$ therefore identifies one concrete waveform in the evaluation set. The trained surrogate predicts, for each candidate, $
U_{p,\epsilon}
$ and $\{V_{O_j,p,\epsilon}\}_{j=1}^{d^2}$,
which are then used to reconstruct the implemented channel and compare it against arbitrary target gates. In this way, the trained graybox becomes a lookup engine for evaluating practical controllability over a finite, hardware-realisable control family.

\paragraph{Reconstruction of the implemented channel}
The $V_{O_j}$ predicted by the graybox model are not themselves quantum channels. The complete collection of predicted expectation values can therefore be used to reconstruct the process matrix associated with the trained graybox model's predicted response. 
For each transformed control $\theta$, the graybox model predicts a coherent propagator and a set of effective observable-space correction operators. 
These are used to reconstruct the implemented channel in a fixed tomography basis.

Let $\{B_i\}_{i=1}^{d^2}$ be a tomographically complete operator basis for the system Hilbert space, and let $\{O_j\}_{j=1}^{d^2}$ be the tomographically complete measurement basis used in the probing protocol.
For each control realisation $(p,\epsilon)$, the surrogate predictions determine expectation values
\begin{align}
b_{ij}(p,\epsilon)
=
\mathrm{Tr}\!\left[
V_{O_j,p,\epsilon}\;
U_{p,\epsilon}\,B_i\,U_{p,\epsilon}^\dagger\;
O_j
\right].
\label{eq:b_ij_method}
\end{align}
Stacking these values produces a vector $\mathbf{b}(p,\epsilon)\in\mathbb{C}^{d^4}$.

These statistics define a linear reconstruction problem,
\begin{align}
\mathbf{b}(p,\epsilon)
=
A\,\mathrm{vec}\!\bigl(J_{p,\epsilon}\bigr),
\end{align}
where $J_{p,\epsilon}$ is the Choi matrix of the realised process and $A$ is a fixed invertible matrix determined by the chosen probing basis. The realised process is reconstructed as
\begin{align}
\mathrm{vec}\!\bigl(J_{p,\epsilon}\bigr)
=
A^{-1}\mathbf{b}(p,\epsilon)\end{align}
The reconstructed matrix is divided by the Hilbert-space dimension \(d \) to obtain the normalised Choi representation used in the fidelity calculation.

For a target gate $G$ with Choi matrix $J_{G}$, the realised process is scored using the Choi fidelity
\begin{align}
F(G;p,\epsilon)
=
\mathrm{Tr}\!\left[
\sqrt{
\sqrt{J_{G}}\,J_{p,\epsilon}\,\sqrt{J_{G}}
}
\right]
\in[0,1].
\label{eq:root_choi_method}
\end{align}
This fidelity is sensitive both to coherent mismatch and to non-unitary open-system distortions, making it appropriate for controllability analysis in the present setting.

\subsection{Gate optimisation over the admissible control family}
\label{sec:ch4_gate_optimization}
For clarity, we first present the construction for a single normalised budget coordinate; the multidimensional case used later in the results follows by evaluating the same optimisation and aggregation procedure over a grid of joint budget values.

To characterise practical controllability statistically rather than for a single target, we draw \(N_G\) target gates
\begin{align}
\{G_g\}_{g=1}^{N_G}\subset \mathrm{SU}(d)
\end{align}
independently from the Haar measure. For each target gate \(G_g\), we evaluate the fidelity over the full sampled control family.

Let \((p,\alpha)\) denote a generic sampled control realisation, where \(p\) indexes the baseline pulse and \(\alpha\) indexes the corresponding constrained variant generated under the chosen control parametrisation.
The resulting fidelity values form a performance table over the sampled control family, $F(G_g;p,\alpha).$
To impose an explicit control budget, we assign to each candidate waveform a normalised resource coordinate
\begin{align}
\alpha_{p,\alpha}\in[0,1].
\end{align}
Given a budget level \(A_0\in(0,1]\), a control realisation is admissible if
\begin{align}
\alpha_{p,\alpha}\le A_0.
\end{align}
The best-achievable fidelity for target gate \(G_g\) at budget level \(A_0\) is then
\begin{align}
F_g^\star(A_0)
:=
\max_{\alpha_{p,\alpha}\le A_0}
F(G_g;p,\alpha),
\label{eq:Fstar_1D_method}
\end{align}
with best-achievable infidelity
\begin{align}
\mathcal{I}_g(A_0):=1-F_g^\star(A_0).
\label{eq:Jg_1D_method}
\end{align}
Thus \(F_g^\star(A_0)\) is the best process fidelity achievable for target gate \(G_g\) under the imposed control budget, while \(\mathcal{I}_g(A_0)\) measures the remaining shortfall from ideal performance.

\subsection{Controllability metrics}

\label{sec:ch4_controllability_metrics}

All controllability summaries in this work are derived from the distribution of best-achievable infidelity over the Haar ensemble. For a fixed budget level \(A_0\), the collection $\{\mathcal{I}_g(A_0)\}_{g=1}^{N_G}$ defines the distribution of best-achievable infidelities over the Haar-random target gates. 
This collection defines the empirical best-achievable-infidelity distribution,
\begin{equation}
    \mu_{A_0}
    =
    \frac{1}{N_G}
    \sum_{g=1}^{N_G}
    \delta_{\mathcal{I}_g(A_0)},
\end{equation}
where $\delta_x$ denotes a point mass at $x$. A controllability summary may then be obtained by applying a chosen functional $\Psi$ to this distribution,
\begin{equation}
    \mathrm{CM}_{\Psi}(A_0)
    =
    \Psi\!\left(\mu_{A_0}\right).
\end{equation}
The choice of $\Psi$ depends on the controllability question of interest and may, for example, describe threshold-based coverage, tail or worst-case behaviour, or the integrated area of the distribution. In the numerical studies reported in this paper, a cutoff-free area functional is used as the primary scalar summary. 
Let
\begin{align}
C_{A_0}(x):=\Pr[\mathcal{I}_g(A_0)\le x]
\end{align}
denote the empirical cumulative distribution function of the best-achievable infidelities at budget level \(A_0\). 
For this choice, we define the area-based controllability metric
\begin{align}
\mathrm{CM}(A_0)= \int_0^{x_{\max}}\bigl[1-C_{A_0}(x)\bigr]\,dx,
\label{eq:CM_method}
\end{align}
where \(x_{\max}\) is chosen as the largest observed best-achievable infidelity across the distributions included in a given comparison and is held fixed across the compared budgets and models. Since \(1-C_{A_0}(x)\) is the survival function of the best-achievable infidelity distribution, \(\mathrm{CM}(A_0)\) measures the total infidelity area remaining away from ideal controllability. Smaller values therefore indicate that the best-achievable fidelities are concentrated closer to unity, while larger values indicate poorer controllability.

For the area functional selected here, since $\mathcal{I}_g(A_0)\geq 0$ and $x_{\max}$ spans the observed support, the survival-function integral is equal to the empirical mean best-achievable infidelity,
\begin{equation}
    \mathrm{CM}(A_0)
    =
    \frac{1}{N_G}
    \sum_{g=1}^{N_G}
    \mathcal{I}_g(A_0).
\end{equation}
Equivalently,
\begin{equation}
    \mathrm{CM}(A_0)
    = 
    W_1\!\left(\mu_{A_0},\delta_0\right),
\end{equation}
where $W_1$ denotes the $1$-Wasserstein distance and $\delta_0$ is the ideal point mass at zero infidelity, corresponding to a distribution in which every target gate has zero best-achievable infidelity \cite{peyre2019computational}.
These identities characterise the particular area functional used in the reported numerical studies rather than the underlying distributional controllability construction.

For clarity, the construction above is written for a single budget coordinate \(A_0\). In the results section, the same gate-wise optimisation and infidelity-aggregation procedure is applied both to one-dimensional budget sweeps and to the joint two-dimensional amplitude--bandwidth cap plane.
At each joint cap pair $(A_{\mathrm{cap}},B_{\mathrm{cap}})$, the same area functional is applied to the corresponding best-achievable-infidelity distribution, so that the resulting $\mathrm{CM}(A_{\mathrm{cap}},B_{\mathrm{cap}})$ surface may likewise be interpreted pointwise as the $1$-Wasserstein distance from that distribution to $\delta_0$.
\section{Results}

\subsection{System model}
In this section, we present the numerical implementation of the proposed controllability framework for the specific case \(d=2\).
The physical device is modelled as a single driven qubit coupled to a finite-dimensional bath mode, with total Hilbert space $
\mathcal{H}=\mathcal{H}_{\mathrm{q}}\otimes \mathcal{H}_{\mathrm{bath}},
$
where $\mathcal{H}_q=\mathbb{C}^2$ and $\mathcal{H}_{bath}=\mathbb{C}^4$.
The simulations were performed over a fixed control window using two control channels corresponding to the \(x\)- and \(y\)-quadratures of the driven qubit. The numerical parameter values used throughout are given in Supplemental Table~S1.

In the rotating frame, the implemented Hamiltonian takes the form
\begin{align}
H(t)
&=
\frac{\delta_{q1}}{2}\,\sigma_z
+
\omega_d\,a^\dagger a
+
\Bigl(
V_{S,1}\,\sigma_+ a
+
V_{S,1}^\ast a^\dagger \sigma_-
\Bigr)
\nonumber\\
&\quad
+
f_{x}(t)\,\sigma_x
+
f_{y}(t)\,\sigma_y
+
\beta_1(t)\,g_1\,\sigma_z,
\label{eq:results_hamiltonian}
\end{align}
where \(\sigma_{x,y,z}\) are the Pauli operators acting on the qubit subsystem, \(a\) and \(a^\dagger\) are the annihilation and creation operators of the bath mode, \(\delta_{q1}=\omega_{q1}-\omega_d\) is the qubit-drive detuning, \(V_{S,1}\) is the qubit-bath coupling strength, and \(\beta_1(t)\) is a classical dephasing noise process coupled through \(g_1\sigma_z\).

The control fields \(f_{x}(t)\) and \(f_{y}(t)\) are generated from Gaussian pulse families on the two quadratures,
\begin{align}
f_{x}(t)
&=
\frac{\Omega_1}{2}
\sum_{n=1}^{n_{\max}}
A_n^{(x)}
\exp\!\left[
-\frac{(t-\tau_n^{(x)})^2}{2\sigma^2}
\right],\\
f_{y}(t)
&=
-\frac{\Omega_1}{2}
\sum_{n=1}^{n_{\max}}
A_n^{(y)}
\exp\!\left[
-\frac{(t-\tau_n^{(y)})^2}{2\sigma^2}
\right],
\end{align} 
where \(n_{\max}=10\), the pulse centres \(\tau_n\) are placed on a fixed equally spaced temporal grid over \([0,T]\), and a common Gaussian width \(\sigma\) is sampled for each pulse realisation.
The open-system dynamics were studied under three cases: a closed-system reference model, a classical-noise model, and a combined quantum-plus-classical noise model.
In the classical-noise case, the stochastic dephasing term $\beta_1(t)$ was
generated using the spectral model defined in
Supplemental Eq.~(S1). The numerical noise parameters and Monte Carlo ensemble size are given in Supplemental Table~S1.

In the combined quantum-plus-classical noise model, dissipative auxiliary-mode dynamics were included using the decay and excitation collapse channels defined in Supplemental Eqs.~(S2) and~(S3), with the corresponding rates given in Supplemental Table~S1.
The numerical propagation follows the procedures introduced in Supplemental Sec.~II. In the closed-system reference, all noise and system--bath coupling terms are set to zero and the nominal controlled dynamics are propagated over the simulation interval. 
In the classical-noise model, each sampled dephasing trajectory defines a deterministic Hamiltonian evolution, and the final responses are averaged over $K$ realisations. In the combined quantum-plus-classical model, the qubit and truncated auxiliary mode are propagated using the vectorised GKSL equation in Supplemental Eq.~(S4) for each sampled classical noise trajectory, after which the final expectation values are averaged over the ensemble.

Three physical settings were considered throughout: a closed-system reference model, a classical-noise model, and a combined quantum-plus-classical noise model. This follows the intended structure of the study, in which the methodology is presented generally at the qudit level, while the results are given for a concrete qubit implementation.
For the numerical study, pulse--response simulations were performed for a library of randomised qubit control pulses. For each example, amplitudes were sampled uniformly within the admissible interval defined by \(A_{\max}(\sigma)\), the Gaussian centres were fixed on the temporal grid, and the common pulse width was sampled between \(\sigma_{\min}\) and \(\sigma_{\max}\). 
The final-time outputs used for graybox training were expectation values of a tomographically complete probing set.
In the qubit implementation, the initial states were chosen as eigenprojectors of the Pauli operators tensored with the bath vacuum, and the measured observables were the Pauli operators acting on the qubit subsystem. The initial-state and measurement selections form a tomographically complete probing set for the qubit process.

The numerical study uses the parameter values in Supplemental Table~S1. 
The width bounds \(\sigma_{\min}\) and \(\sigma_{\max}\) define the sampling interval for the common Gaussian width used in each pulse realisation, while \(A_{\max}(\sigma)\) gives the corresponding width-dependent amplitude bound.

\subsection{Graybox model training}
A separate graybox model was trained for each of the three physical regimes. In each case, the raw pulse parameters were normalised to \([0,1]\) before training. Amplitudes were rescaled relative to the width-dependent maximum amplitude \(A_{\max}(\sigma)\), temporal centres were normalised by \(T\), and widths were normalised linearly between \(\sigma_{\min}\) and \(\sigma_{\max}\). 
For the one-qubit implementation, the resulting input representation was a tensor in $\mathbb{R}^{N_{\mathrm{ex}}\times n_{\max}\times 6}$, where $N_{\mathrm{ex}}$ denotes the number of pulse examples in each physical-regime dataset. For each Gaussian component, the six input features were the amplitude and temporal centre for each of the $x$- and $y$-quadrature control channels, with the common pulse width repeated in each channel representation.
In each case, the graybox model received the normalised pulse tensor as input and was trained to predict the vector of measured expectation values associated with the probing basis. After training, the model was evaluated over the scaled pulse families considered in this study, and the resulting predictions were then used in the surrogate-based tomography and fidelity pipeline described below.

Training and validation mean-squared-error histories for the three graybox response models are provided in Supplemental Fig.~S1.

\begin{figure*}[htbp]
    \centering

    \includegraphics[width=0.98\textwidth]
    {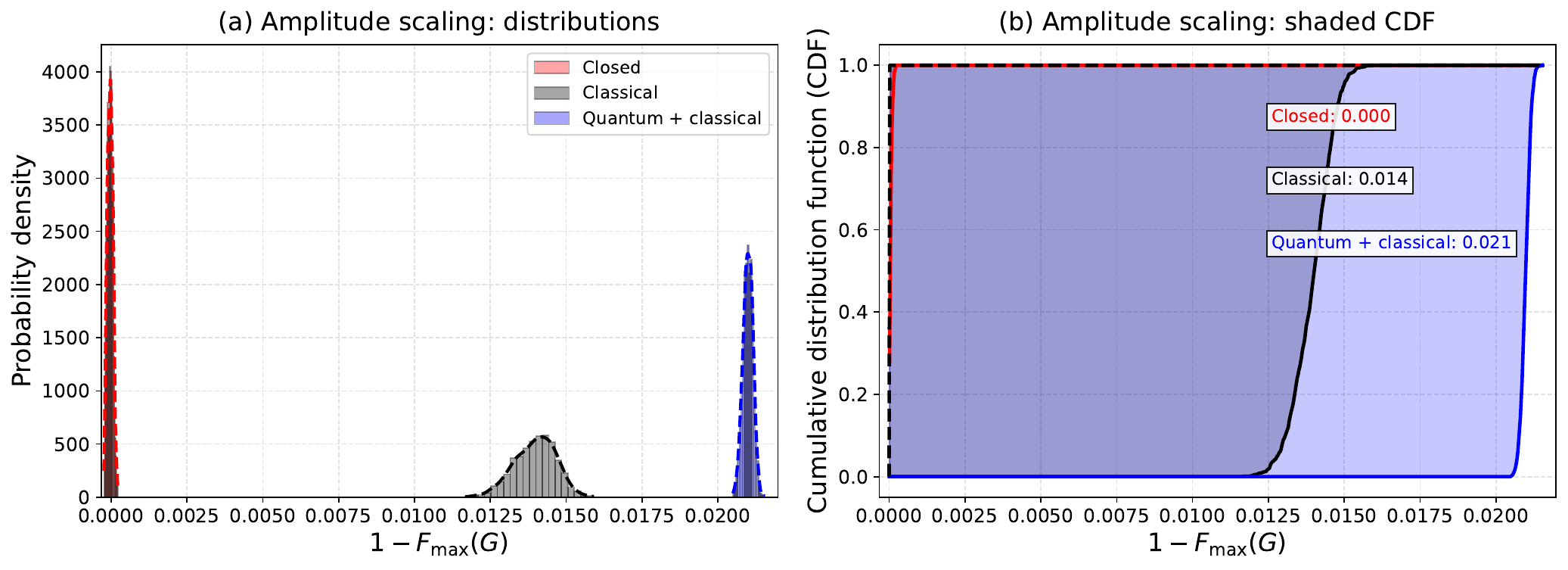}

    \vspace{3mm}

    \includegraphics[width=0.98\textwidth]
    {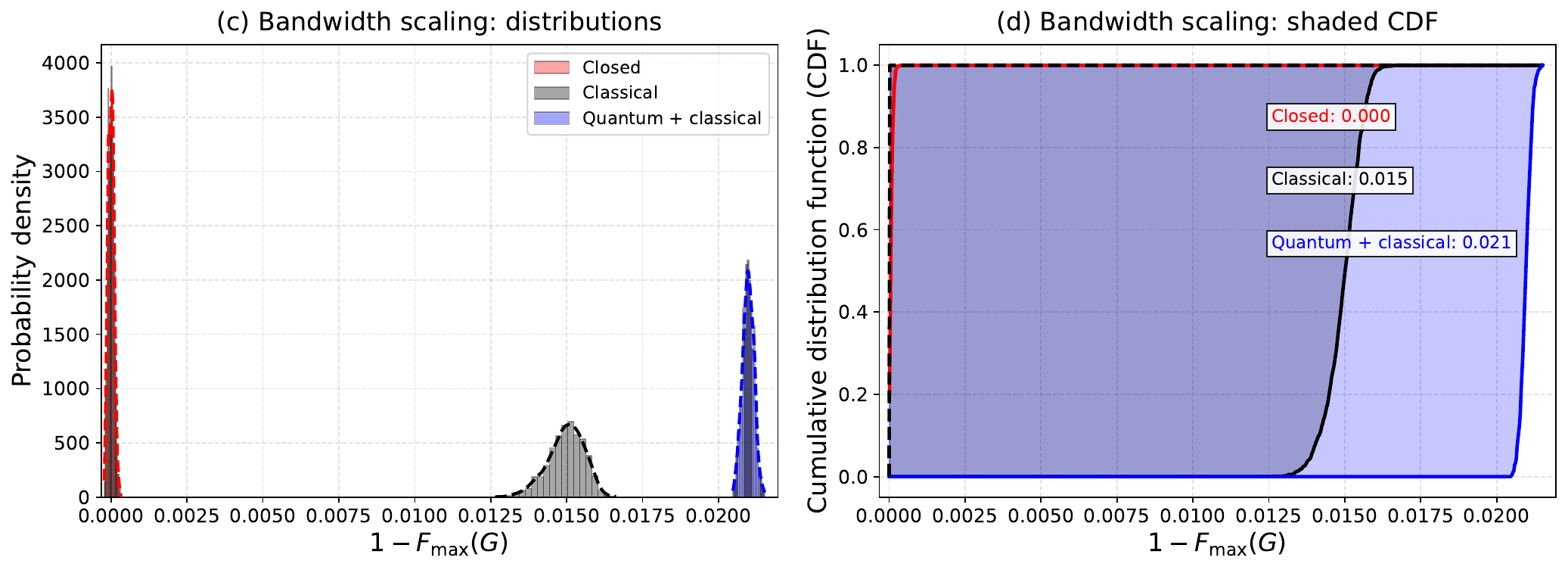}

    \caption{\textbf{One-dimensional resource-bounded controllability studies.}
    Distributions and empirical cumulative distribution functions of the best
    achievable infidelity
    \(1-F_{\max}(G)\) over the Haar-random target-gate ensemble.
    The amplitude study varies the waveform amplitude resource, while the
    Gaussian-width study varies the common temporal width \(\sigma\), which
    acts as the bandwidth-related resource coordinate. Results are shown for
    the closed-system, classical-noise, and combined
    quantum-plus-classical-noise regimes. The shaded regions in the CDF panels
    represent the area used to calculate the corresponding controllability
    measure.}
    \label{fig:1d_slices}
\end{figure*}
\subsection{Pulse scaling}

With the graybox model fixed, the next step was to generate post-training pulse families over which controllability could be evaluated.
This subsection instantiates the admissible-control-family construction introduced in Section~\ref{sec:ch4_gate_optimization} for the present qubit implementation.
At the implementation level, the normalised budget coordinates are tied to physical pulse resources rather than abstract deformation parameters. 
The amplitude and bandwidth variations used below are defined at the level of the parametrised pulse family. Amplitude scaling is a uniform multiplicative rescaling of the waveform and therefore preserves the time-domain pulse shape, pulse centres, widths, and relative quadrature structure. Bandwidth scaling is implemented through the Gaussian width parameter and therefore changes the temporal envelope by construction. In the joint study, each baseline pulse is expanded over a Cartesian amplitude--bandwidth grid and then filtered by simultaneous amplitude and bandwidth bounds.
For amplitude scaling, each baseline waveform \(f_p(t)\) is first converted to physical units and assigned a peak amplitude
\begin{align}
A_p := \max_t |f_p(t)|.
\end{align}
A global admissible amplitude cap \(A_{\max,\mathrm{global}}\) is then fixed from the pulse family, and the normalised amplitude coordinate determines a target cap through
\begin{align}
A_{\mathrm{cap}}(\alpha_{\mathrm{amp}})= \alpha_{\mathrm{amp}}A_{\max,\mathrm{global}}.
\end{align}
Each baseline pulse is then rescaled by a pulse-dependent factor so that its transformed waveform reaches this target level. Writing the transformed waveform as
\begin{align}
f_{p,\epsilon}(t)=\epsilon_p\,f_p(t),
\end{align}
the corresponding scaling factor is
\begin{align}
\epsilon_p=\frac{A_{\mathrm{cap}}(\alpha_{\mathrm{amp}})}{A_p}=\frac{\alpha_{\mathrm{amp}}A_{\max,\mathrm{global}}}{A_p}.
\end{align}
For the one-dimensional amplitude sweep, the coordinate \(\alpha_{\mathrm{amp}}\in[0,1]\) was sampled on an \(E=200\)-point grid. In the implemented qubit setting, the global amplitude limit was taken to be
\begin{align}
A_{\max,\mathrm{global}} =\frac{\pi}{\sqrt{2\pi}\,\sigma_{\min}}.
\end{align}
The widths and pulse centres were held fixed along this slice.
For bandwidth scaling, the Gaussian pulse width \(\sigma\) is used as the underlying physical control parameter, with effective bandwidth identified through the inverse relation
\begin{align}
BW \propto \frac{1}{\sigma}.
\end{align}
The minimum and maximum admissible bandwidths are therefore induced by the width limits,
\begin{align}
BW_{\min}=\frac{1}{\sigma_{\max}},
\qquad BW_{\max}=\frac{1}{\sigma_{\min}},
\end{align}
and the normalised bandwidth coordinate determines a target bandwidth level through
\begin{align}
BW(\alpha_{\mathrm{bw}})
=(1-\alpha_{\mathrm{bw}})BW_{\min}+\alpha_{\mathrm{bw}}BW_{\max}.
\end{align}
The corresponding target pulse width is
\begin{align}
\sigma_{\mathrm{cap}}(\alpha_{\mathrm{bw}}) =\frac{1}{BW(\alpha_{\mathrm{bw}})}.
\end{align}
Each baseline pulse is then reassigned a width through a pulse-dependent scaling of its baseline width \(\sigma_p\),
\begin{align}
\sigma_{p,\epsilon}=\epsilon_p\,\sigma_p,
\end{align}
where
\begin{align}
\epsilon_p =\frac{\sigma_{\mathrm{cap}}(\alpha_{\mathrm{bw}})}{\sigma_p}.
\end{align}
The same baseline testing library was expanded in three ways: a one-dimensional amplitude scaling family, a one-dimensional bandwidth scaling family, and a joint two-dimensional amplitude--bandwidth family.
For the one-dimensional bandwidth sweep, the pulse widths were reassigned directly according to a global normalised bandwidth coordinate
\begin{align}
\alpha_{\mathrm{bw}}\in[0,1],
\end{align}
sampled on a grid of \(N=200\) points. 
The corresponding physical bandwidth was varied linearly between
\(BW_{\min}\) and \(BW_{\max}\), while the normalised amplitude
coordinates and pulse centres were held fixed. Under the
width-dependent amplitude normalisation, this preserves the area of
each Gaussian component.
For the full joint trade-off study, each baseline pulse was expanded onto a Cartesian grid
\begin{align}
(\alpha_{\mathrm{amp}},\alpha_{\mathrm{bw}})\in[0,1]\times[0,1]
\label{eq:joint_ab_grid}
\end{align}
with \(N_{\alpha_{amp}}=21\) amplitude points and \(N_{\alpha_{bw}}=21\) bandwidth points. 

Along the amplitude branch, each baseline waveform was rescaled toward the global amplitude cap \(A_{\max,\mathrm{global}}\).
Along the bandwidth branch, the widths were reassigned according to the linear bandwidth interpolation between \(BW_{\min}\) and \(BW_{\max}\). 
This produced, for each baseline pulse, a two-dimensional family of amplitude--bandwidth variants indexed by the sampled grid values.
The one-dimensional amplitude and bandwidth sweeps are therefore the corresponding one-parameter controllability analyses, while the joint heatmaps are obtained from the separate two-dimensional amplitude--bandwidth construction.



\begin{figure*}[t]
    \centering
    \begin{subfigure}[t]{0.5\textwidth}
        \centering
        \includegraphics[width=\linewidth]{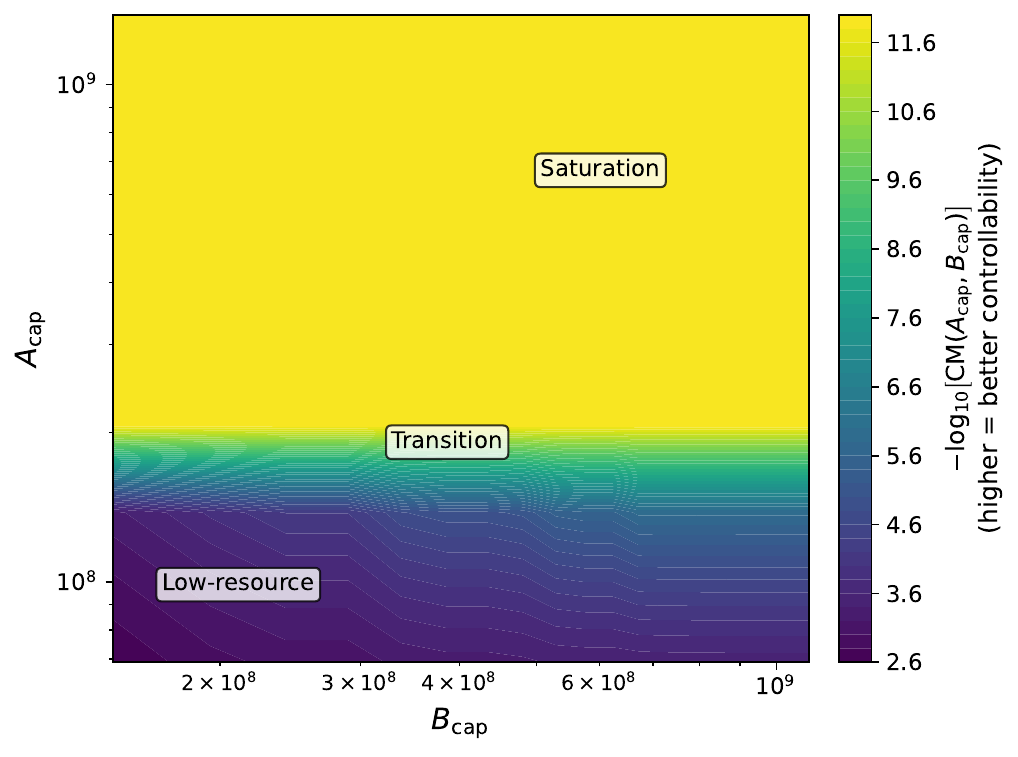}
        \caption{Closed-system reference model.}
        \label{fig:f3a_closed}
    \end{subfigure}\hfill
    \begin{subfigure}[t]{0.5\textwidth}
        \centering
        \includegraphics[width=\linewidth]{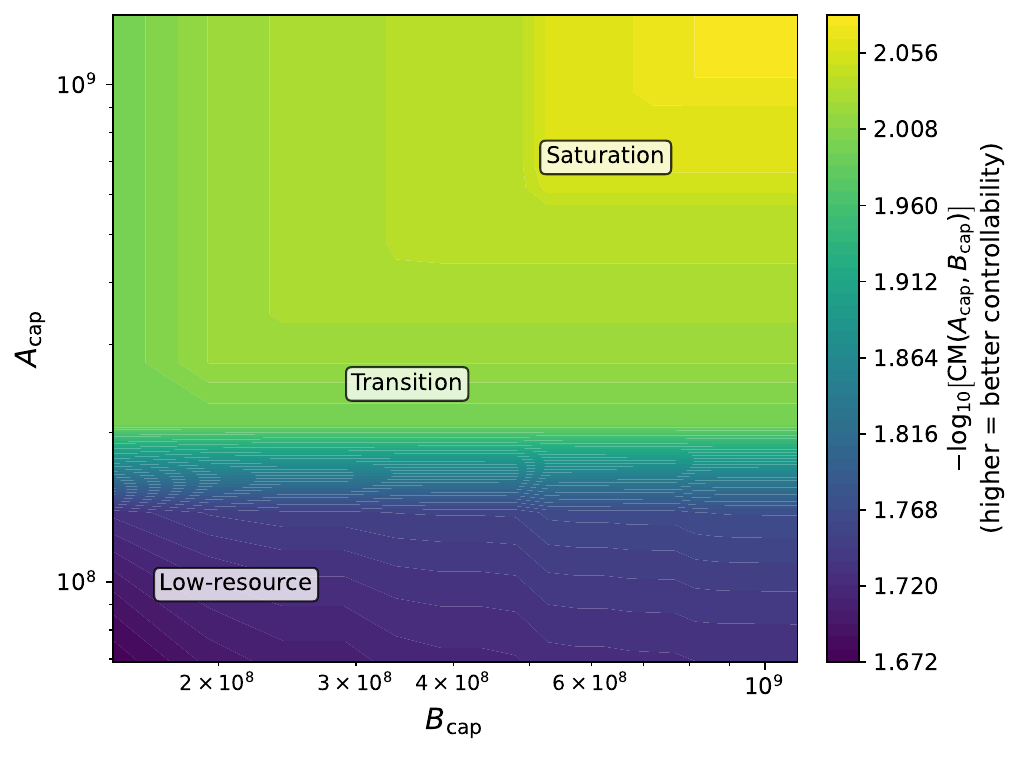}
        \caption{Classical-noise model.}
        \label{fig:f3b_weaknoise}
    \end{subfigure}

    \vspace{1mm}

    \begin{subfigure}[t]{0.5\textwidth}
        \centering
        \includegraphics[width=\linewidth]{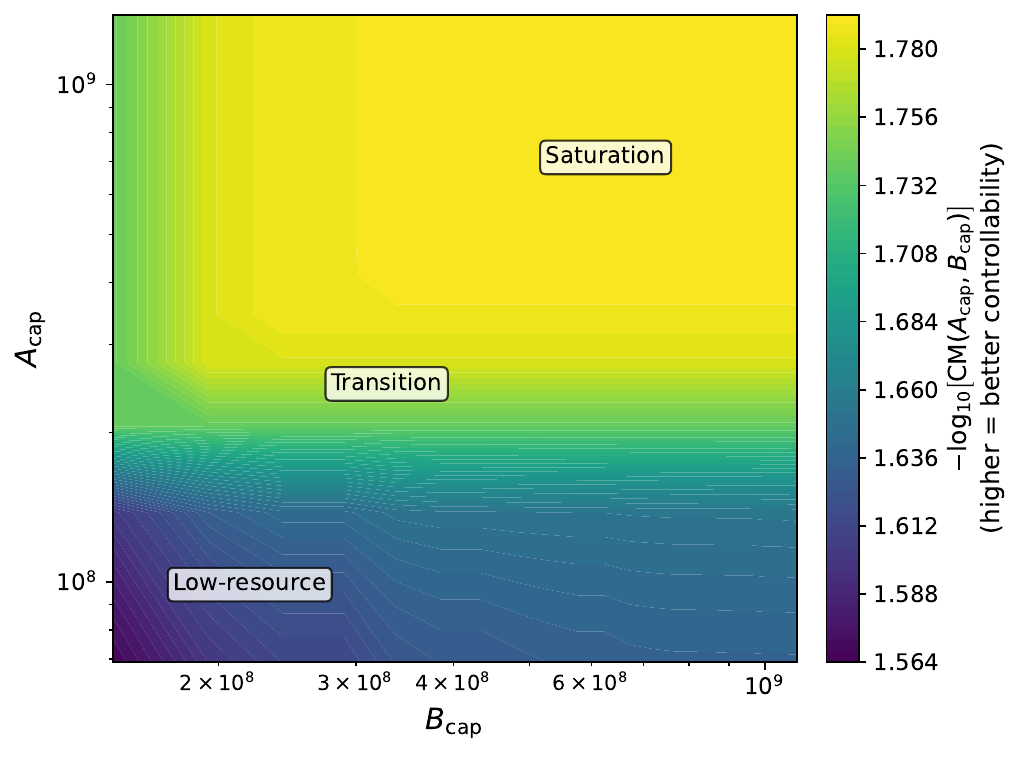}
        \caption{Quantum-plus-classical noise model.}
        \label{fig:f3c_strongnoise}
    \end{subfigure}\hfill

    \caption{\textbf{Two-dimensional controllability surfaces under joint amplitude and bandwidth caps.} For each cap pair $(A_{\mathrm{cap}},B_{\mathrm{cap}})$, the admissible pulse family is first defined by the feasible-set condition from the methodology. For each Haar-random target gate, the best feasible fidelity is then computed over that admissible family, converted to a best-achievable infidelity, and aggregated into the area-based controllability metric $\mathrm{CM}(A_{\mathrm{cap}},B_{\mathrm{cap}})$. The plotted quantity is $-\log_{10}\!\bigl(\mathrm{CM}(A_{\mathrm{cap}},B_{\mathrm{cap}})\bigr)$, so larger values indicate improved controllability. The annotations indicate qualitative resource regimes: low-resource, where performance is resource-limited; transition, where controllability improves rapidly as the caps are relaxed; and saturation, where further resource yields diminishing improvement. Panels show the same construction for (a) the closed-system reference model, (b) the classical-noise  model, and (c) the quantum-plus-classical noise model.}
    \label{fig:2d_cm_surfaces}
\end{figure*}

\subsection{Numerical controllability results}
\paragraph*{One-dimensional controllability studies:}

For the one-dimensional amplitude and bandwidth studies, the best-achievable infidelity was evaluated for every Haar-random target gate over the admissible subset of the corresponding scaled pulse family. Figure ~\ref{fig:1d_slices} reports the one-dimensional amplitude and bandwidth studies.
The histogram panels report the empirical distribution of best achievable infidelity, $1-F_{max}(G)$, in the full-budget case, and the CDF panels report the corresponding cumulative distributions for the three physical regimes.
The shaded area between each empirical CDF and the ideal step function gives the one-dimensional area-based controllability metric.
\paragraph*{Two-dimensional controllability studies:}
Figure~\ref{fig:2d_cm_surfaces} reports the joint amplitude--bandwidth controllability surfaces for the three physical regimes considered in this work.
For the joint study, the normalised cap coordinates
\((A_0,B_0)\in[0,1]^2\) are mapped to the physical cap values
\begin{align}
A_{\mathrm{cap}}=A_0 A_{\max,\mathrm{global}},
\end{align}
\begin{align}
B_{\mathrm{cap}}=BW_{\min}+B_0(BW_{\max}-BW_{\min}).
\end{align}
For each sampled realisation, the corresponding physical requirement measures are
\begin{align}
A_{\mathrm{req}}(p,e_a,e_b)
&=
\max_t \bigl|f_{p,e_a,e_b}(t)\bigr|,
\\
BW_{\mathrm{req}}(p,e_a,e_b)
&=
\frac{1}{\sigma^{\mathrm{eff}}_{p,e_a,e_b}},
\end{align}
where \(e_a\in\{1,\dots,N_{\alpha_{amp}}\}\) and \(e_b\in\{1,\dots,N_{\alpha_{bw}}\}\) label the sampled amplitude and bandwidth variants generated from the Cartesian grid in Eq.~\eqref{eq:joint_ab_grid}, and \(\sigma^{\mathrm{eff}}_{p,e_a,e_b}\) denotes the effective realized pulse width for that family member.
The feasible set associated with a given cap pair is then
\begin{align}
\mathcal{F}(A_{\mathrm{cap}},B_{\mathrm{cap}})
&=
\Bigl\{
(p,e_a,e_b):
A_{\mathrm{req}}(p,e_a,e_b)\le A_{\mathrm{cap}},
\nonumber\\
&\hspace{2.8em}
BW_{\mathrm{req}}(p,e_a,e_b)\le B_{\mathrm{cap}}
\Bigr\}.
\label{eq:feasible_set_method}
\end{align}
For each Haar-random target gate \(G_g\), the best feasible fidelity inside a given cap cell is
\begin{align}
F_g^\star(A_{\mathrm{cap}},B_{\mathrm{cap}})
:=
\max_{(p,e_a,e_b)\in\mathcal{F}(A_{\mathrm{cap}},B_{\mathrm{cap}})}
F_g(p,e_a,e_b),
\end{align}
with corresponding best-achievable infidelity
\begin{align}
\mathcal{I}_g(A_{\mathrm{cap}},B_{\mathrm{cap}})
:=
1-F_g^\star(A_{\mathrm{cap}},B_{\mathrm{cap}}).
\end{align}
Applying the area-based controllability metric from Section~\ref{sec:ch4_controllability_metrics} to the resulting Haar-ensemble distribution yields the two-dimensional surface \(\mathrm{CM}(A_{\mathrm{cap}},B_{\mathrm{cap}})\). In each panel, the plotted quantity is \(-\log_{10}\!\bigl(\mathrm{CM}(A_{\mathrm{cap}},B_{\mathrm{cap}})\bigr)\), and the same construction is used in all three cases under identical amplitude--bandwidth cap coordinates.

\section{Discussion}
The practical question posed at the outset of this work was not whether the device is controllable in a theoretical Lie-algebraic or binary sense, but rather which target operations remain achievable to high fidelity under a fixed hardware-realisable control family and explicit resource constraints, and how that capability changes as additional amplitude and bandwidth become available.
In the present framework, that question is answered through a controllability surface defined under explicit amplitude and bandwidth limits $\mathrm{CM({A_\mathrm{cap}},{B_\mathrm{cap}})}$,
which summarises the distribution of gate-wise best-achievable infidelity over the Haar ensemble for each pair of amplitude and bandwidth limits.
The figures herein therefore convert practical open-system controllability into a directly interpretable resource landscape over the joint amplitude-bandwidth plane.

The one-dimensional controllability results provide a direct view of how best-achievable performance is distributed over the Haar ensemble.
In the CDF panels of Fig.~\ref{fig:1d_slices}, a steeper curve concentrated near zero best-achievable infidelity indicates that a larger fraction of target gates can be implemented ideally, or close to ideally, whereas a broader or right-shifted curve indicates that best-achievable performance remains farther from the ideal limit across more of the Haar ensemble.
In the ideal case, corresponding to the closed system, this distribution would collapse to a step function at zero best-achievable infidelity, corresponding to an area-based controllability metric \(\mathrm{CM}=0\).
The one-dimensional results therefore retain some of the intuition of a binary controllability statement while also quantifying how far the physical device remains from the idealised limit.
These one-dimensional results also show a consistent ordering across the three physical regimes. In the amplitude-scaling CDF in Fig.~\ref{fig:1d_slices}b, the annotated \(\mathrm{CM}\) values for the closed-system, classical-noise, and quantum-plus-classical-noise regimes are approximately \(0.000\), \(0.014\), and \(0.021\), respectively, while the corresponding bandwidth-scaling values in Fig.~\ref{fig:1d_slices}d are approximately \(0.000\), \(0.015\), and \(0.021\). The closed-system case is therefore effectively ideal in both studies, while the larger classical-noise \(\mathrm{CM}\) value in the bandwidth study indicates poorer best-achievable performance along that resource slice.
The two-dimensional controllability results shown in Fig.~\ref{fig:2d_cm_surfaces} indicate that the joint amplitude--bandwidth plane exhibits a clear low-resource--transition--saturation structure in all three noise regimes.
In the part of the plane where both \(A_{\mathrm{cap}}\) and \(B_{\mathrm{cap}}\) are small, the feasible pulse family is too restricted to support broad gate-set controllability, so the plotted quantity \(-\log_{10}\!\bigl(\mathrm{CM}(A_{\mathrm{cap}},B_{\mathrm{cap}})\bigr)\) remains low.
As the amplitude and bandwidth limits are relaxed, the surface enters a transition region in which controllability improves more rapidly as the control family expands, approaching a higher-performance regime in which further increases yield smaller gains.
This transition is most visible in the noisy cases, where the useful-control region occupies a smaller portion of the displayed plane. 
Overall, the closed-system case shows the strongest controllability, the classical-noise case is degraded, and the quantum-plus-classical-noise case is the most restricted.
At the same time, noise does not simply lower the controllability level at fixed control; it also deforms the useful region itself.
In the closed-system case, a broad portion of the displayed amplitude--bandwidth plane quickly enters the high-performance regime once moderate amplitude is available. In the classical-noise case, the rise toward that regime is shifted toward larger amplitude and bandwidth limits relative to the closed-system reference.
In the quantum-plus-classical-noise case, the overall controllability level is reduced further, but the surface is also reshaped rather than simply shifted uniformly: bandwidth dependence becomes more concentrated in the low-\(B_{\mathrm{cap}}\) region, even though the absolute \(-\log_{10}(\mathrm{CM})\) values remain below those of the classical-noise case.

A key feature of the present construction is that the joint controllability surface is parameterised directly in terms of physical amplitude and bandwidth limits rather than purely abstract deformation coordinates. The normalised cap coordinates \((A_0,B_0)\) are mapped to the physical cap variables \((A_{\mathrm{cap}},B_{\mathrm{cap}})\) within the chosen pulse family and evaluation pipeline.
The physical caps therefore do not merely indicate qualitative trends, but show how much control resource must be made available to enter the useful high-performance regime and where further increases begin to yield only marginal improvement.

The one-dimensional studies and the two-dimensional joint amplitude–bandwidth studies suggest a clear resource hierarchy. At small amplitude caps, increasing bandwidth alone yields only modest gains because the admissible family remains too weak to realise a broad class of target gates, whereas increasing amplitude produces the sharper initial rise and more clearly controls entry into the useful-control regime. In the present implementation, amplitude therefore acts as the primary resource governing the onset of controllability, while bandwidth acts as a secondary but still necessary resource that enlarges and refines the regime once sufficient drive strength is available.

This hierarchy is consistent with the structure of the sampled pulse family. Within each pulse realisation, the amplitudes of the individual Gaussian components are varied independently, whereas a common width $\sigma$ is shared across all components. The training data therefore span a substantially richer set of amplitude configurations than width configurations. Consequently, the surrogate is trained over a higher-dimensional set of amplitude configurations, while the bandwidth-related coordinate is restricted to variation of the common Gaussian width. This difference in the sampled control family provides a plausible explanation for the sharper controllability transition observed along the amplitude coordinate.
The resulting surfaces are therefore useful not only for identifying the amplitude and bandwidth levels required to enter a useful high-performance regime and where further resource yields only marginal gains, but also for diagnosing whether the chosen pulse family is disproportionately constrained along one control direction.

More generally, the complete procedure can be repeated under a different amplitude--bandwidth budget, noise regime, admissible pulse family, or experimental control configuration to produce a second empirical best-achievable-infidelity distribution. The \(1\)-Wasserstein distance between the two distributions then quantifies the corresponding change in the overall distribution of achievable gate performance.
This provides a basis for the control-family benchmarking indicated in Fig.~\ref{fig:cm_workflow}. In a laboratory setting, the framework could be applied while varying parameters that either directly define the programmed control pulse or indirectly modify the waveform implemented at the device through downstream elements of the control chain. Examples include the pulse parametrisation, amplitude and bandwidth limits, filtering, calibration settings, and hardware transfer characteristics. Repeating the gate-wise optimisation for each configuration would produce a controllability distribution for each case, allowing their effects on achievable performance across the target-gate ensemble to be compared quantitatively.

In this sense, these results do more than compare fidelities across models: they provide an operational map of the control-resource cost required to recover high-quality gate synthesis as the physical environment becomes increasingly realistic. 

The current study has several limitations. The reported surfaces are computed for the chosen driven-system model, pulse parametrisation, and sampled control ranges. The graybox model is only evaluated within the pulse family used for training and post-training analysis, so extrapolation outside this region is not assessed. Bandwidth is represented through the Gaussian width parameter rather than through an explicit hardware filter or transfer function. The controllability metrics are also computed over finite pulse libraries and finite Haar-random target samples, and therefore provide numerical resource-dependent summaries rather than exact analytic controllability criteria. In deployment to a physical system, the validity of the graybox model would also be time-limited because drift in the system dynamics, control electronics, or measurement response could progressively invalidate the learned correction, requiring periodic validation or retraining.

\noindent\textbf{Data and code availability} The source code and data generated in this study are available upon  request from the corresponding author.\\

\noindent\textbf{Acknowledgments} YM acknowledges support through the Australian Government Research Training Program Scholarship. AP acknowledges an RMIT University Vice-Chancellor’s Senior Research Fellowship and a Google Faculty Research Award. This work was supported by the Australian Government through the Australian Research Council under the Centre of Excellence scheme (No: CE170100012). This research was also undertaken with the assistance of resources from the National Computational Infrastructure (NCI Australia), an NCRIS enabled capability supported by the Australian Government. \\

\noindent\textbf{Author Contributions}
All authors contributed to the writing of the paper. YM developed the resource-bounded controllability framework, conducted all qubit numerical experiments, and performed the associated analyses, with inputs from AY and AP.
AY provided technical and methodological guidance. 
AP provided technical advice and critical feedback on the manuscript.
AY and AP supervised the project.

\bibliographystyle{apsrev4-2}
\bibliography{references}

\end{document}


\title{Supplemental Material for Resource-bounded controllability benchmarking of open quantum systems}

\author{Yule Mayevsky}
\affiliation{Quantum Photonics Laboratory and Centre for Quantum Computation and Communication Technology, RMIT University, Melbourne, VIC 3000, Australia}

\author{Akram Youssry}
\affiliation{Quantum Photonics Laboratory and Centre for Quantum Computation and Communication Technology, RMIT University, Melbourne, VIC 3000, Australia}
\affiliation{School of Electrical Engineering and Telecommunications, UNSW Sydney, Sydney, NSW 2052, Australia}

\author{Alberto Peruzzo}
\affiliation{Quantum Photonics Laboratory and Centre for Quantum Computation and Communication Technology, RMIT University, Melbourne, VIC 3000, Australia}
\affiliation{Quandela, Massy, France}

\maketitle

\section{System and numerical parameters}

\begin{table}[htbp]
\caption{System and numerical parameters used for the resource-bounded controllability
simulations.}
\label{tab:supp_simulation_parameters}
\centering
\small
\renewcommand{\arraystretch}{1.15}

\begin{tabularx}{\textwidth}{
    @{}
    >{\raggedright\arraybackslash}p{0.42\textwidth}
    >{\centering\arraybackslash}X
    >{\centering\arraybackslash}X
    >{\centering\arraybackslash}X
    @{}
}
\toprule

\multirow{2}{*}{\textbf{Parameter}}
&
\multicolumn{3}{c}{\textbf{Common simulation parameters}}
\\
\cmidrule(lr){2-4}

&
\multicolumn{3}{c}{\textbf{Value}}
\\
\midrule

System dimension \(d\)
&
\multicolumn{3}{c}{\(2\)}
\\

Evolution time \(T\)
&
\multicolumn{3}{c}{\(0.25\,\mu\mathrm{s}\)}
\\

Time-grid points \(M\)
&
\multicolumn{3}{c}{\(1000\)}
\\

Classical-noise realisations \(K\)
&
\multicolumn{3}{c}{\(1000\) }
\\

Auxiliary-mode dimension \(d_{\mathrm{bath}}\)
&
\multicolumn{3}{c}{\(4\)}
\\

Initial auxiliary-mode state
&
\multicolumn{3}{c}{Vacuum state \(\lvert0\rangle\langle0\rvert\)}
\\

Control channels \(n_{\mathrm{ctrl}}\)
&
\multicolumn{3}{c}{\(2\)}
\\

Gaussian components per quadrature \(n_{\max}\)
&
\multicolumn{3}{c}{\(10\)}
\\

Qubit frequency \(\omega_{q1}\)
&
\multicolumn{3}{c}{\(2\pi\times5.33\,\mathrm{GHz}\)}
\\

Drive frequency \(\omega_d\)
&
\multicolumn{3}{c}{\(2\pi\times5.30\,\mathrm{GHz}\)}
\\

Detuning \(\delta_{q1}\)
&
\multicolumn{3}{c}{\(2\pi\times30\,\mathrm{MHz}\)}
\\

Control scaling \(\Omega_1\)
&
\multicolumn{3}{c}{\(2\)}
\\

Minimum Gaussian width \(\sigma_{\min}\)
&
\multicolumn{3}{c}{\(0.04\,T/(n_{\max}+1)\)}
\\

Maximum Gaussian width \(\sigma_{\max}\)
&
\multicolumn{3}{c}{\(0.30\,T/(n_{\max}+1)\)}
\\

Maximum Gaussian coefficient \(A_{\max}(\sigma)\)
&
\multicolumn{3}{c}{\(\pi/(\sqrt{2\pi}\sigma)\)}
\\

Low-frequency PSD coefficient
&
\multicolumn{3}{c}{\(10^{9}\)}
\\

High-frequency PSD coefficient
&
\multicolumn{3}{c}{\(10^{-9}\)}
\\

Dataset size, $N_{ex}$
&
\multicolumn{3}{c}{\(10^{4}\) pulse--response examples per regime}
\\

\midrule

\multirow{2}{*}{\textbf{Parameter}}
&
\multicolumn{3}{c}{\textbf{Physical-regime parameters}}
\\
\cmidrule(lr){2-4}

&
\textbf{Closed}
&
\textbf{Classical}
&
\textbf{\shortstack{Quantum +\\classical}}
\\
\midrule

Classical dephasing coupling \(g_1\)
&
\(0\)
&
\(50\)
&
\(50\)
\\

Qubit--bath coupling \(V_{S,1}\)
&
\(0\)
&
\(0\)
&
\(2\pi\times45\,\mathrm{MHz}\)
\\

Auxiliary-mode decay rate \(\gamma_0\)
&
\(0\)
&
\(0\)
&
\(2\pi\times1\,\mathrm{kHz}\)
\\

Auxiliary-mode excitation rate \(\gamma_1\)
&
\(0\)
&
\(0\)
&
\(0\)
\\

\bottomrule
\end{tabularx}
\end{table}

\section{Noise models and numerical simulation}
\label{sec:numerical_simulations}

The simulation interval \([0,T]\) was divided into \(M\) uniform intervals of width
\begin{equation*}
\Delta t=\frac{T}{M}.
\end{equation*}
Time-dependent quantities were evaluated at
\begin{equation*}
t_k
=
\left(k-\frac{1}{2}\right)\Delta t,
\qquad
k=1,\ldots,M.
\end{equation*}

\subsection{Classical-noise simulation}

For each control waveform, \(K\) independent realisations
\begin{equation*}
\left\{
\beta_1^{(r)}(t)
\right\}_{r=1}^{K}
\end{equation*}
of the classical dephasing process were generated. For the \(r\)-th realisation, the Hamiltonian \(H^{(r)}(t)\) was obtained from the Hamiltonian in the main text by replacing \(\beta_1(t)\) with \(\beta_1^{(r)}(t)\). In the classical-noise regime,
\(V_{S,1}=\gamma_0=\gamma_1=0\).

Coloured classical noise was generated by constructing a frequency-domain representation and transforming it into the time domain. The finite sampling window gives
\begin{equation*}
\Delta f=\frac{1}{T},
\qquad
f_{\mathrm{Nyq}}=\frac{1}{2\Delta t}.
\end{equation*}
Random phases were applied to amplitudes proportional to \(\sqrt{S_{\beta}(f)}\), followed by an inverse Fourier transform. The real component was retained as the noise trajectory \(\beta_1^{(r)}(t_k)\).
The classical coloured-noise model had the spectral form 
\begin{equation}
S_{\beta}(f) = \frac{\alpha_1}{f} +
\alpha_2 f,
\label{eq:classical_noise_psd}
\end{equation}
where \(\alpha_1\) controls the low-frequency \(1/f\) contribution and \(\alpha_2\) controls the high-frequency contribution. The singular point at \(f=0\) was regularised numerically. The values of \(\alpha_1\) and \(\alpha_2\) are given in Supplemental Table~S1.

At each time step, the sampled Hamiltonian was
\begin{equation*}
H_k^{(r)}
=
H^{(r)}(t_k),
\end{equation*}
with corresponding short-time propagator
\begin{equation*}
U_k^{(r)}
=
\exp\!\left[
-iH_k^{(r)}\Delta t
\right].
\end{equation*}
The final propagator for the \(r\)-th realisation was approximated by
the ordered product
\begin{equation*}
U^{(r)}(T)
\approx
U_M^{(r)}
U_{M-1}^{(r)}
\cdots
U_1^{(r)}.
\end{equation*}

For initial qubit state \(\rho_i(0)\) and measured observable \(O_j\),
the response for one noise realisation was
\begin{equation*}
E_{ij}^{(r)}
=
\mathrm{Tr}\!\left[
U^{(r)}(T)
\rho_i(0)
U^{(r)\dagger}(T)
O_j
\right].
\end{equation*}
The ensemble-averaged response was estimated by
\begin{equation*}
E_{ij}
\approx
\frac{1}{K}
\sum_{r=1}^{K}
E_{ij}^{(r)}.
\end{equation*}

\subsection{Combined quantum-plus-classical-noise simulation}

In the combined quantum-plus-classical-noise regime, each sampled trajectory \(\beta_1^{(r)}(t)\) was inserted into the Hamiltonian in the main text. The qubit and truncated auxiliary mode were then propagated as a joint system on
\begin{equation*}
\mathcal{H}_{\mathrm q}
\otimes
\mathcal{H}_{\mathrm{bath}}.
\end{equation*}
For initial qubit state \(\rho_i(0)\), the initial joint state was
\begin{equation*}
\rho_i^{(r)}(0)
=
\rho_i(0)
\otimes
\lvert0\rangle\langle0\rvert,
\end{equation*}
where the auxiliary mode was initialised in the vacuum state.

The auxiliary-mode decay and excitation collapse operators were
\begin{equation}
L_0
=
\sqrt{\gamma_0}
\left(
I_{\mathrm q}\otimes a
\right),
\label{eq:aux_mode_decay_collapse}
\end{equation}
and
\begin{equation}
L_1
=
\sqrt{\gamma_1}
\left(
I_{\mathrm q}\otimes a^\dagger
\right),
\label{eq:aux_mode_excitation_collapse}
\end{equation}
where \(I_{\mathrm q}\) is the identity on the qubit Hilbert space and \(a\) is the annihilation operator of the retained auxiliary mode. The values of \(\gamma_0\), \(\gamma_1\), and \(d_{\mathrm{bath}}\) are
given in Supplemental Table~S1. 
For each classical-noise realisation, the joint density operator was
propagated in vectorised form. Let \begin{equation*}
\hat{\rho}_i^{(r)}(t) = \operatorname{Vec}\!\left[ \rho_i^{(r)}(t) \right].
\end{equation*}
The vectorised master equation was
\begin{equation}
\begin{aligned}
\dot{\hat{\rho}}_i^{(r)}(t)
&=
\mathcal{M}^{(r)}(t)
\hat{\rho}_i^{(r)}(t),
\\
\mathcal{M}^{(r)}(t)
&=
-i
\left[
I\otimes H^{(r)}(t)
-
H^{(r)\mathsf T}(t)\otimes I
\right]
\\
&\quad
+
\sum_{\mu=0}^{1}
\Bigg[
L_\mu^{*}\otimes L_\mu
-
\frac{1}{2}
I\otimes L_\mu^\dagger L_\mu
-
\frac{1}{2}
\left(
L_\mu^\dagger L_\mu
\right)^{\mathsf T}
\otimes I
\Bigg],
\end{aligned}
\label{eq:liouville-gksl}
\end{equation}
where \(I\) is the identity on the joint qubit--auxiliary-mode Hilbert space. At each time step, the short-time Liouville-space propagator was
\begin{equation*}
\mathcal{E}_k^{(r)}
=
\exp\!\left[
\mathcal{M}^{(r)}(t_k)\Delta t
\right].
\end{equation*}
The final vectorised density operator was approximated by
\begin{equation*}
\hat{\rho}_i^{(r)}(T)
\approx
\mathcal{E}_M^{(r)}
\mathcal{E}_{M-1}^{(r)}
\cdots
\mathcal{E}_1^{(r)}
\hat{\rho}_i^{(r)}(0).
\end{equation*}

After propagation,
\(\hat{\rho}_i^{(r)}(T)\) was reshaped into the joint density operator \(\rho_i^{(r)}(T)\). The response for the \(r\)-th realisation was
\begin{equation*}
E_{ij}^{(r)}
=
\mathrm{Tr}\!\left[
\rho_i^{(r)}(T)
\left(
O_j\otimes I_{\mathrm{bath}}
\right)
\right],
\end{equation*}
where \(I_{\mathrm{bath}}\) is the identity on the truncated auxiliary mode. The final response was obtained by averaging over the
classical-noise realisations,
\begin{equation*}
E_{ij}
\approx
\frac{1}{K}
\sum_{r=1}^{K}
E_{ij}^{(r)}.
\end{equation*}

The closed-system reference, the classical-noise and combined quantum-plus-classical-noise parameters used for dataset generation are reported in the Supplemental Table~S1.
\clearpage

\section{Graybox model training diagnostics}
\begin{figure}[htbp]
    \centering
    \includegraphics[
        width=0.82\textwidth,
        height=0.8\textheight,
        keepaspectratio
    ]{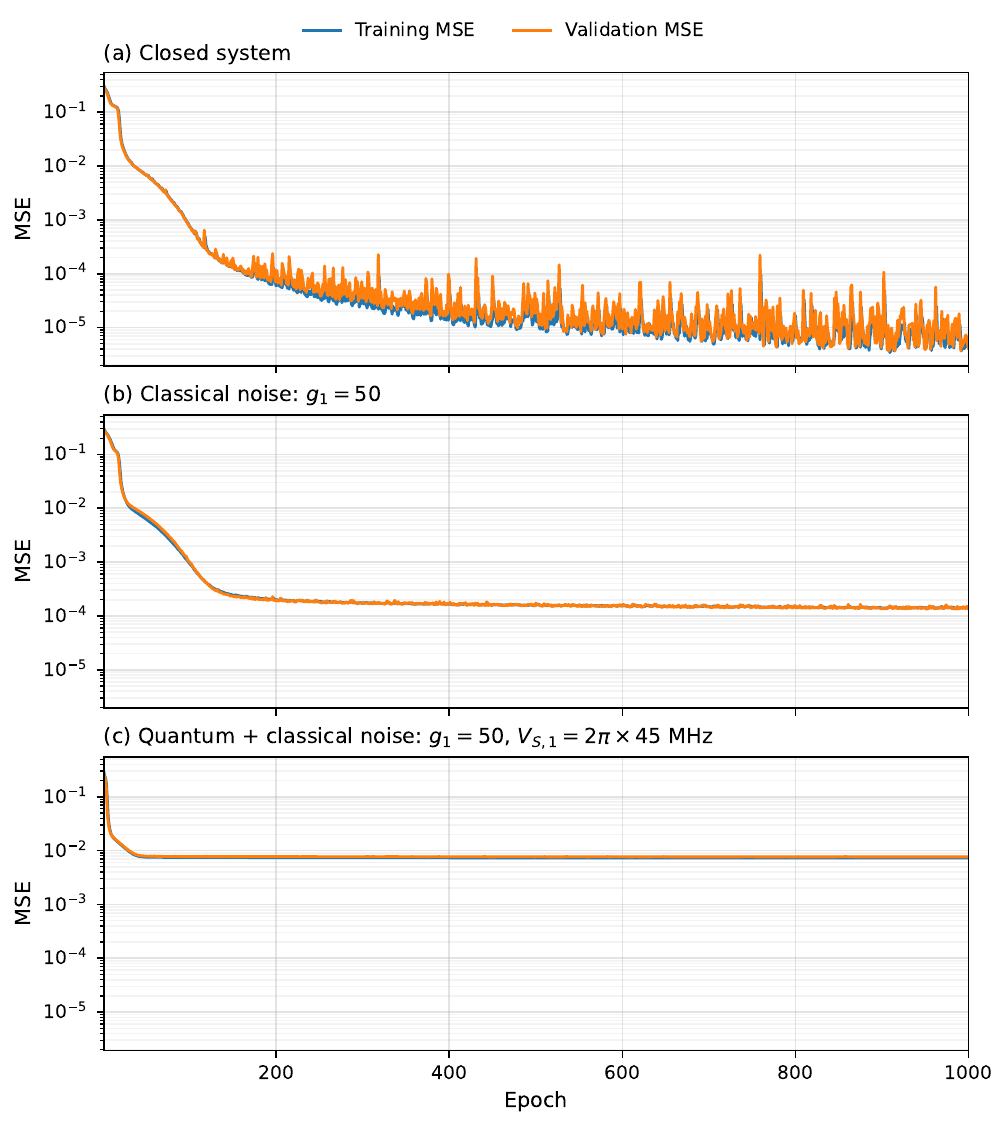}
    \caption{\textbf{Training and validation mean-squared-error histories for the graybox response models used in this work:}
    The panels correspond to (a) the closed-system model, (b) the
    classical-noise model with \(g_1=50\), and (c) the combined
    quantum-plus-classical noise model with \(g_1=50\) and
    \(V_{S,1}=2\pi\times45\,\mathrm{MHz}\). The vertical axis is logarithmic.}
    \label{fig:supp_graybox_training_curves}
\end{figure}